\documentstyle[11pt]{article} \textwidth 15cm
\begin{document}

\centerline{\bf\large SEARCH FOR AMBIENT NEUTRALINO DARK MATTER AT
ACCELERATOR}

\vspace{1cm}

\centerline{Tai-Fu Feng$^{1,2}$, Xue-Qian Li$^{1,2}$, Wen-Gan Ma$^{1,3}$ and
X. Zhang$^{1,4}$}

\vspace{0.7cm}

1 CCAST (World Laboratory) P.O. Box 8730, 100080, Beijing, China.\\

2 Department of Physics, Nankai University, Tianjin, 300070, China.\\

3 Department of Modern Physics, University of Science and Technology
of China,\\

\hspace{1cm} Hefei, 230026, China\\

4 Institute of High Energy Physics, P.O. Box 918-4, 100039, Beijing, China.

\vspace{2cm}

\begin{center}
\begin{minipage}{11cm}
\noindent Abstract

We investigate the possibility of using  accelarator beam particles to
collide with the ambient neutralino
dark matter particles in cosmic rays as a way to search
for the cold dark matter.
We study in detail its
inelastic and elastic scattering with the projectile
particles at electron-positron colliders and discuss the possible
experimental signals and the relevent background.
\end{minipage}
\end{center}

\vspace{2cm}

\vspace{0.2cm}

There is strong evidence for the existence of a large amount of cold dark
matter (CDM) in the universe.
The leading candidate of the cold dark matter
particles in the minimal supersymmetric standard model
with R parity  is neutralino \cite{s1,Brh}.
The possibility of sneutrino being a
candidate of CDM is ruled out unless it as heavy as 17 TeV which is
unfavorable in all  resonable theories\cite{Falk,Hall}.

In the traditional method of directly searching for
such CDM SUSY particles, the velocity of the SUSY CDM particles is very
small \cite{Cowsik}, their kinetic energy is not sufficient to cause any
inelastic scattering, so that a clear observation would be difficult due to
existence of a flood of background trajectories.
In this paper we propose a new
detection mechanism of neutralino cold dark matter by using the
accelarator beam particles to collide with the ambient neutralino dark
matter particles.
Namely we use the projectile particles produced
in accelerator ($e^{\pm},p(\bar p)$ or even $\gamma$) to be incident on the
dark matter particles. Because most of the cosmic ray particles interact
with ordinary matter via electromagnetic processes, they cannot penetrate into
the tunnel with high vacuum, but the dark matter particles  only
participate in "weak" interaction and can come into the tunnel.
Concretely, we let
the beam (say, $e^-$) go through a highly vacuumized tunnel while the
colliding
beam ($e^+$) is turned off. Once the projectile electrons bombard on
neutralinos ($\tilde{\chi}_1^0$),
because of the
large available energy of $e^-$, inelastic processes may occur as
\begin{equation}
\tilde{\chi}_1^0+e^-\rightarrow\tilde{\chi}_1^-+\nu_e.
\end{equation}

 The charged SUSY particles
$\tilde{\chi}_1^-$
would make trajectories in detector with magnetic
field.
Because they are much heavier than proton and pion etc. ordinary particles,
 they can be
identified easily from background products. There are also elastic processes
\begin{equation}
\tilde{\chi}_1^0+e^-\rightarrow \tilde{\chi}^0_{1}+e^-,\;\;\;\;
{\nu}_e+e^-,
\end{equation}
where the projectile electron declines from its beam direction. However, this
process might be smeared with the background effect such as
$$n+e^-\rightarrow n+e^-,$$
where $n$ is the nucleon left in the tunnel, even though it is highly
vacuumized.
 We will discuss the background problem again later in this letter.

The density of cold dark matter in our ambient universe has been estimated
\cite{Data}. There have also been many theoretical works concerning the
SUSY relic density after the Big Bang \cite{Sred}. Thus we can immediately
evaluate the flux of the SUSY cold dark matter \cite{Li}. The observed event
number of the suggested reaction can be obtained as
\begin{equation}
N=\rho_{DM}\cdot\rho_{beam}\cdot |\vec v_{rel}|\sigma S\cdot l\cdot t=
\rho_{DM}\cdot {1\over 2} L\sigma S\cdot l\cdot t,
\end{equation}
where $\rho_{DM},\rho_{beam}$ are the densities of dark matter and beam
particles respectively, $|\vec v_{rel}|$ is their relative velocity, $\sigma$
is the cross section of the reaction, $S$ is the cross section of the beam,
$l$ is the length of the available detection region, $t$ is the time duration
for measurement. Since the velocity of the coming SUSY particles is much
lower than that of the beam particle, $\rho_{beam}|\vec v_{rel}|\sim L/2$
where $L$ is the named luminosity.

In 1972, a peculiar event of heavy cosmic ray particle was observed in the
cloudy chamber of the Yunan Cosmic Ray Station (YCRS) \cite{Wuli}. Recently,
the event was re-analyzed \cite{Ho} and it is identified as that a heavy
neutral particle $C^0$ came in and bombarded on a proton to produce a
heavy charged particle $C^+$ as well as a proton and  $\pi^-$. Their analysis
confirmed that the mass of the heavy neutral cosmic ray particle $C^0$ is
greater than 43 GeV and the mass difference
$$\Delta M=M_{C^+}-M_{C^0}<0.270\;{\rm GeV}.$$

If taking this result seriously, one would be tempted to conclude that the
coming neutral $C^0$ is a SUSY dark matter particle $\tilde{\chi}_1^0$
and the produced heavy charged particle is
$\tilde{\chi}_1^+$
accordingly. In this case, the available energy of the
Beijing Electron-Positron Collider (BEPC) is sufficient to cause an
inelastic scattering where the projectile particle $e^-$ of 2 GeV hits
the coming $\tilde{\chi}_1^0$
to produce $\tilde{\chi}_1^-$.
If the mass difference of $\tilde{\chi}^{\pm}$ and $\tilde{\chi}^0$
is as large as a few tens GeV,
the
BEPC energy is not sufficient, and one
needs
to invoke machines with higher energy, such as, LEP or hadron colliders.
In this paper we will take
 $\Delta M<1$ GeV for a detail study, and then we will briefly give a
general discussion on the case for larger $\Delta M$.

In calculation of the cross section $\sigma$ in eqs.(1), (2) and (3)
without losing generality we assume that  only one generation sfermion is
light
and one can  neglect the sfermion mixing among different generations.
The mass matrices of neutralinos and charginos
can be found in Ref\cite{s1}. For the mixing of the first generation slepton (right
and left fields), we have
\begin{equation}
M_{\tilde{e}}^{2}=\left(
\begin{array}{cc}
-\frac{e^{2}(\upsilon_{1}^{2}-\upsilon_{2}^{2})(1-2c_{w}^{2})}{8s_{w}c_{w}}
+M_{e}^{2}+M_{L}^{2} & \frac{1}{\sqrt{2}}\bigg(
\sqrt{2}M_{e}\mu+\upsilon_{1}h_{s_{l}}^{1}\bigg) \\
\frac{1}{\sqrt{2}}\bigg(\sqrt{2}M_{e}\mu+\upsilon_{1}h_{s_{l}}^{1}\bigg)&
\frac{e^{2}(\upsilon_{1}^{2}-\upsilon_{2}^{2})}{4c_{w}^{2}}+M_{e}^{2}
+M_{R}^{2}
\end{array}
\right).
\label{matr1}
\end{equation}
The mixing matrix $Z_{\tilde{e}}$ is defined as
\begin{equation}
\label{mix}
Z_{\tilde{e}}^{\dagger}M_{\tilde{e}}^{2}Z_{\tilde{e}}=diag(M_{\tilde{e}_{1}}
^{2}, M_{\tilde{e}_{2}}^{2}).
\end{equation}
The mass of the electron-sneutrino is:
\begin{equation}
M_{\tilde{\nu}_{e}}^{2}=M_{L}^{2}-\frac{e^{2}(\upsilon_{1}^{2}-\upsilon_{
2}^{2})}{8s_{w}^{2}c_{w}^{2}}.
\end{equation}

At tree level, the squared mass, $M_{H_{1}^{0}}^{2}$, of the light Higgs
boson has an upper bound which is given by $M_{Z}^{2}\cos^{2}2\beta$. This
is already
below the experimental lower bound of LEP2\cite{s2}. However, radiative
corrections can raise the upper bound on $M_{H_{1}^{0}}^{2}$
dramatically\cite{s3}. The dominant contribution is
\begin{equation}
\bigtriangleup M_{H_{1}^{0}}^{2}=\frac{3M_{t}^{4}}{2\pi^{2}\upsilon^{2}}
\ln\frac{M_{\tilde{t}}^{2}}{M_{t}^{2}} ,
\end{equation}
where $M_{t}$ is the top quark mass and $M_{\tilde{t}}$ the top-squark mass.
In our numerical calculation, we take the correction into account.

When the
kinematics is permissive, several inelastic reactions such as
$e^{-}+\tilde{\chi}_{1}^{0}
\rightarrow \tilde{\nu}_{e} + W^{-}(H^{-}_{1})$, $e^{-}+\tilde{\chi}_{1}^{0}
\rightarrow \tilde{e}_{i}^{-} + Z^{0}(H^{0}, A^{0})$ $(i=1$, $2)$
etc. can occur. For the moment we consider only
inelastic channels $e^{-}+\tilde{\chi}_1^0
\rightarrow \tilde{\chi}_1^- + \nu_{e}$ and as well the elastic processes
$e^{-}+\tilde{\chi}_{1}^{0}
\rightarrow  e^{-}+\tilde{\chi}_{1}^{0}$.
For the channel
$e^{-}+\tilde{\chi}_{1}^{0}\rightarrow \tilde{\chi}_{1}^{-} + \nu_{e}$,
the amplitudes are given as
\begin{eqnarray}
&&{\cal M}_{\tilde{e}_{i}^{-}}^{s}=-\frac{i}{s-M_{\tilde{e}_{i}^{-}}^{2}}
\sum\limits_{\sigma_{1}\sigma_{2}}A_{\sigma_{1}}^{(1)}B_{\sigma_{2}}^{(1)}
\bar{\upsilon}(p_{2})\omega_{\sigma_{1}}u(p_{1})\bar{u}(p_{4})
\omega_{\sigma_{2}}\upsilon(p_{3}), \nonumber \\
&&{\cal M}_{W}^{u}=-\frac{i}{u-M_{W}^{2}}\frac{e^{2}}{2s_{w}^{2}}
\sum\limits_{\sigma}C_{\sigma}^{(1)}\upsilon^{T}(p_{3})C^{-1}
\gamma_{\mu}\omega_{-}u(p_{1})
\bar{u}(p_{4})\gamma^{\mu}\omega_{\sigma}C\bar{\upsilon}^{T}
(p_{2}) \nonumber \\
&&\hspace{1.5cm}=-\frac{i}{u-M_{W}^{2}}\frac{e^{2}}{2s_{w}^{2}}
\sum\limits_{\sigma}C_{\sigma}^{(1)}\bar{u}(p_{3})\gamma_{\mu}
\omega_{-}u(p_{1})
\bar{u}(p_{4})\gamma^{\mu}\omega_{\sigma}u(p_{2}), \nonumber \\
&&{\cal M}_{\tilde{\nu}_{e}}^{t}=\frac{-i}{t-M_{\tilde{\nu}_{e}}^{2}}
\sum\limits_{\sigma_{1}\sigma_{2}}D_{\sigma_{1}}^{(1)}E_{\sigma_{2}}^{(2)}
\bar{\upsilon}(p_{2})\omega_{\sigma_{1}}\upsilon(p_{3})\bar{u}(p_{4})
\omega_{\sigma_{2}}u(p_{1}),
\label{amp1}
\end{eqnarray}
where the couplings are
\begin{eqnarray}
&&A_{-}^{(1)}=\frac{e}{\sqrt{2}s_{w}c_{w}}Z_{\tilde{e}}^{1i}(Z_{N}^{11}s_{w}
+Z_{N}^{21}c_{w}), \nonumber \\
&&A_{+}^{(1)}=-\frac{\sqrt{2}e}{c_{w}}Z_{\tilde{e}}^{2i}Z_{N}^{11*},
\nonumber \\
&&B_{-}^{(1)}=0, \nonumber \\
&&B_{+}^{(1)}=-\frac{e}{s_{w}}Z_{\tilde{e}}^{1i*}Z_{-}^{1j*}, \nonumber \\
&&C_{-}^{(1)}=Z_{N}^{21}Z_{+}^{1j*}-\frac{Z_{N}^{41}Z_{+}^{2j*}}{\sqrt{2}},
\nonumber \\
&&C_{+}^{(1)}=Z_{N}^{21*}Z_{-}^{1j}+\frac{Z_{N}^{3i*}Z_{-}^{2j}}{\sqrt{
2}},\nonumber \\
&&D_{-}^{(1)}=\frac{e}{\sqrt{2}s_{w}c_{w}}(Z_{N}^{11}s_{w}-Z_{N}^{21}c_{w}),
\nonumber \\
&&D_{+}^{(1)}=0, \nonumber \\
&&E_{-}^{(1)}=-\frac{e}{s_{w}}Z_{+}^{1j}, \nonumber \\
&&E_{+}^{(1)}=0.
\label{coupling1}
\end{eqnarray}
$Z_{N}$, $Z_{+}$, $Z_{-}$ are defined in \cite{s1} and $Z_{\tilde{e}}$ is given
in (\ref{mix}).

The amplitudes for $e^{-}+\tilde{\chi}_{1}^{0} \rightarrow e^{-} + \tilde{\chi}_{i}^{0}$
(when i=1, it is the elastic scattering case) are
\begin{eqnarray}
&&{\cal M}_{\tilde{e}_{j}^{-}}^{s}=-\frac{i}{s-M_{\tilde{e}_{j}^{-}}^{2}}
\sum\limits_{\sigma_{1}\sigma_{2}}A_{\sigma_{1}}^{(2)}B_{\sigma_{2}}^{(2)}
\bar{\upsilon}(p_{2})\omega_{\sigma_{1}}u(p_{1})\bar{u}(p_{3})\omega_{
\sigma_{2}}\upsilon (p_{4}), \nonumber \\
&&{\cal M}_{Z^{0}}^{t}=\frac{i}{t-M_{Z}^{2}}\sum\limits_{\sigma_{1}
\sigma_{2}}C_{\sigma_{1}}^{(2)}D_{\sigma_{2}}^{(2)}\bar{u}(p_{3})\gamma_{\mu}
\omega_{\sigma_{1}}u(p_{1})\bar{\upsilon}(p_{2})\gamma^{\mu}\omega_{\sigma_{
2}}\upsilon (p_{4}), \nonumber \\
&&{\cal M}_{\tilde{e}_{j}^{-}}^{u}=-\frac{i}{u-M_{\tilde{e}_{j}^{-}}^{2}}
\sum\limits_{\sigma_{1}\sigma_{2}}E_{\sigma_{1}}^{(2)}F_{\sigma_{2}}^{(2)}
\bar{u}(p_{3})\omega_{\sigma_{1}}u(p_{2})\bar{u}(p_{4})\omega_{\sigma_{2}}
u(p_{1}).
\label{amp-2}
\end{eqnarray}
The couplings are written as
\begin{eqnarray}
&&A_{-}^{(2)}=\frac{e}{\sqrt{2}s_{w}c_{w}}Z_{\tilde{e}}^{1j}(Z_{N}^{11}s_{w}
+Z_{N}^{21}c_{w}), \nonumber \\
&&A_{+}^{(2)}=-\frac{\sqrt{2}e}{c_{w}}Z_{\tilde{e}}^{2j}Z_{N}^{11*},
\nonumber\\
&&B_{-}^{(2)}=-\frac{\sqrt{2}e}{c_{w}}Z_{\tilde{e}}^{2j*}Z_{N}^{1i},
\nonumber \\
&&B_{+}^{(2)}=\frac{e}{\sqrt{2}s_{w}c_{w}}Z_{\tilde{e}}^{1j*}
(Z_{N}^{1i*}s_{w}+Z_{N}^{2i*}c_{w}), \nonumber \\
&&C_{-}^{(2)}=\frac{e}{s_{w}c_{w}}(\frac{1}{2}-s_{w}^{2}), \nonumber \\
&&C_{+}^{(2)}=-\frac{es_{w}}{c_{w}}, \nonumber \\
&&D_{-}^{(2)}=\frac{e}{2s_{w}c_{w}}Z_{N}^{41*}Z_{N}^{4i}, \nonumber \\
&&D_{+}^{(2)}=\frac{e}{2s_{w}c_{w}}Z_{N}^{31}Z_{N}^{3i*}, \nonumber \\
&&E_{-}^{(2)}=-\frac{\sqrt{2}e}{c_{w}}Z_{\tilde{e}}^{2j*}Z_{N}^{11},
\nonumber \\
&&E_{+}^{(2)}=\frac{e}{\sqrt{2}s_{w}c_{w}}Z_{\tilde{e}}^{1j}(Z_{N}^{11*}s_{w}
+Z_{N}^{21*}c_{w}), \nonumber \\
&&F_{-}^{(2)}=\frac{e}{\sqrt{2}s_{w}c_{w}}Z_{\tilde{e}}^{1j*}(Z_{N}^{1i}s_{w}
+Z_{N}^{2i}c_{w}), \nonumber \\
&&F_{+}^{(2)}=-\frac{\sqrt{2}e}{c_{w}}Z_{\tilde{e}}^{2j}Z_{N}^{1i*}.
\label{coupling2}
\end{eqnarray}

With the amplitude, we can easily obtain the cross sections by integrating over
the phase space of final states. It is noted that we carry out all the calculations
in the laboratory frame because the velocity of the heavy dark matter particles
is very small compared to that of the projectile beam particles.\\

In our numerical calculations, we take
 $\alpha=1/128.8, M_Z=91.12$ GeV,
$M_W=80.22$ GeV and first assume that
$M_{\tilde{\chi}^-}-M_{\tilde{\chi}^0_1}$
is about 1 GeV.
We will also present the results for
larger mass difference later.
For $e^-+\tilde{\chi}_1^0\rightarrow\tilde{\chi}_1^-+\nu_e$ we take the
SUSY parameters as $\tan\beta=5,M_{\tilde{\chi}^-_2}=200$ GeV.
In the propagators we choose the typical
values for the SUSY particle pole masses as $M_{\tilde{e}_1}
=110$ GeV $M_{\tilde{e}_2}=200$ GeV, $M_{\tilde{\nu}_e}=110$ GeV.
For the Higgs masses, we have $M_{H_1^0}=110$ GeV, $M_{H_2^0}=300$ GeV
and $M_{A^0}=110$ GeV which are commonly adopted in literatures.

In the computations, we consider three possible masses of
$\tilde{\chi}_1^0 $ as
40 GeV, 50 GeV, and 80 GeV.

We find that the cross section of
$e^-+\tilde{\chi}_1^0\rightarrow\tilde{\chi}_1^-+\nu_e$
is of order 100$\mu$b,
which is the typical value for the weak interactions.
We tabulate event numbers of inelastic process $e^-+\tilde{\chi}_1^0\rightarrow
\tilde{\chi}_1^-+\nu_e$ for luminosities and energies of various $e^+e^-$
colliders in Table 1.

\vspace{0.5cm}

\begin{center}
\begin{tabular}{|c|c|c|c|c|c|c|c|}
\hline
& $c-\tau$ factory & BEPC & VEPP-4M & CESR & KEKB & PEP-II & LEP \\
\hline
\parbox[t]{2cm}{
$M_{\tilde{\chi}_1^0}=40$ \\$M_{\tilde{\chi}_1^-}=41$ } &
800 & 20 & 250 & 392 & 375 & 850 & 4950 \\
\hline
\parbox[t]{2cm}{
$M_{\tilde{\chi}_1^0}=50$ \\$M_{\tilde{\chi}_1^-}=51$ } &
1400 & 40  & 380  & 538  & 625  & 1410 & 1300 \\
\hline
\parbox[t]{2cm}{
$M_{\tilde{\chi}_1^0}=80$ \\$M_{\tilde{\chi}_1^-}=81$ } &
16350 & 490  & 11830  & 18676  & 51875  & 32032 & 26850  \\
\hline
\parbox[t]{2cm}{
$M_{\tilde{\chi}_1^0}=40$ \\$M_{\tilde{\chi}_1^-}=60$ } &
$-$ & $-$ & $-$ & $-$ & $-$ & $-$ & 2106 \\
\hline
\parbox[t]{2cm}{
$M_{\tilde{\chi}_1^0}=40$ \\$M_{\tilde{\chi}_1^-}=90$ } &
$-$ & $-$ & $-$ & $-$ & $-$ & $-$ & 2711 \\
\hline
\end{tabular}
\end{center}

\vspace{0.5cm}

\centerline{Table 1}

\vspace{0.3cm}

The estimated event number for
$e^-+\tilde{\chi}_1^0\rightarrow\tilde{\chi}_1^-+\nu_e$. Here the length
of detection region $l$ is taken as 1 m and the masses are in GeV.
In the third row, the number is
unreasonably large and it is because of the pole effect of the propagator and
related to the chosen parameters. The last two rows correspond to larger
$\Delta M=20,\; 50$ GeV, and only the LEP beam energy is capable of making
such inelastic process.
\\

For the elastic processes $e^-+\tilde{\chi}_1^0\rightarrow\tilde{\chi}_1^0
+e^-$,
we evaluated the cross sections in terms of the
parameters introduced above and obtain that the event number of such reactions
at the BEPC luminosity and energy is still about 50 per year.

Experimentally we expect to observe
inelastic scattering between neutralino and the beam particles at $e^+e^-$
machines.
Because all SUSY particles are much heavier than ordinary SM particles, the
trajectories of charged SUSY and SM particles can be easily distinguished
in a strong magnetic field.
At lower beam energies, such as BEPC, the kinetic
energy of the produced chargino is relatively small, so that one
can easily detect them.

The third row of Table 1 show very large number of events, the reason is
due to the pole of
the propagator. But unless the matching happens to occur this way, it is not the case. So we
do not take such large numbers seriously, of course, if the beam energy can be 
adjusted freely in a wide range, the threshold effects might manifest, but the possibility
in practice is slim.

Now let us turn to study the elastic processes where the background may contaminate the
situation. The observation is based on measuring the electrons
scattered from the SUSY
dark matter particles in $e^-+\tilde{\chi}_1^0
\rightarrow e^-+\tilde{\chi}_1^0$.
The background is from
the electrons scattered
from nucleons of the remnant atmosphere in the vacuumized tunnel $e^-+n\rightarrow e^-+n$.
At lower energies, the cross section of scattering can be easily computed and the amplitude is
\begin{equation}
{\cal M}={G_F\over 2\sqrt 2}\bar n\gamma_{\mu}[(-1+{4\over 3}\sin^2\theta_W)+\gamma_5]n
\bar e[(-1+4\sin^2\theta_W)+\gamma_5]e.
\end{equation}
Then we can obtain the cross section. At the same length, the background events are at least
1000 times larger  than the expected events at $1.0315 10^{-6} pa$.
However, the situation can be remedied by careful
analysis on kinematics. 

The maximal differential cross sections of
either $e^-+\tilde{\chi}_1^0\rightarrow
e^-+\tilde{\chi}_1^0$
or $e^-+n\rightarrow e^-+n$ occur near $\theta=\pi/2$.
We have also the energy of the scattered
electron $E'$ as
\begin{equation}
E'={EM\over E(1-\cos\theta)+M},
\end{equation}
where $E$ is the beam energy and in the expression the mass of electron is
ignored, $M$ is the mass of either the SUSY particle or nucleon.
Obviously, if $M\gg E$ we have $E'\sim E$, but
as $E\sim M$, at $\theta=\pi/2$, $E'=ME/(M+E)$. In the BEPC case,
if the electron is scattered
from a SUSY particle, $E'\sim 2$ GeV,
but when it is from the background nucleon, $E'\sim 0.6$
GeV. Therefore, if we set a reasonable cut at, say, 1 GeV for
the kinetic energy of the scattered electron, we can
effectively eliminate the background (the case for $\theta=0$
vicinity should be excluded). However, if the beam energy is
much greater than $M$, for example at LEP and
future colliders, the energy difference  of electron scattered from a heavy
SUSY particle or a nucleon is negligible, so that one cannot
distinguish the source of scattering
and the expected events would be drowned in the ocean of background. 
Therefore, observation of
SUSY dark matter via elastic scattering can only be
feasible at low energy accelerator, and
the energy of BEPC or charm-tau factory is ideal. Moreover,
in that case, the mass difference
of the neutral and charged SUSY particles has no effects at all.

It is noted that the main background is due to scattering between the beam
electron and atmospheric neutron, since only neutron is neutral in the remanat
atmosphere. In fact, the collision between beam electron and other charged particles
in the atmosphere such as proton and electron can be easily  determined by existence
of trajectories of the bounced charged particles and the scattered beam electron.
Therefore our signal can be clearly singled out from the background.
\\

In summary,
we have proposed a new method of detection of the cold dark
matter and
analyzed the possibility of observing inelastic scattering processes
between
beam particles and the ambient neutralino cold dark matter particles.
 We also investigate the situation for elastic scattering.
Our results suggest that if the mass gap between the lightest SUSY particles
$\tilde{\chi}_1^0$
 and their corresponding charged
SUSY particles $\tilde{\chi}_1^-$
is less than 1 GeV,
the BEPC energy is enough to cause inelastic processes where the products
can be easily identified in present experimental facilities.
For large mass gap,
one needs to invoke larger machines such as LEP etc.

Besides we also discuss the possibility of measuring elastic scattering and find
that by carefully setting a kinematic cut for the scattered electron energy, it
is possible to identify if they are scattered by the incident SUSY
particles from large background at BEPC, but not at accelerators with higher
energies.

Our conclusion is that using beam particles to bombard on the coming SUSY
particles may cause inelastic and elastic processes whose products are
measurable and can be identified.

We should point out that Errede and Luk in Ref.\cite{fl} have outlined a
scheme for searching
for terrestrial dark matter at Tevatron. However they have not
studied in detail the processes of the beam particles colliding with 
heavy cold dark matter particles, such as the neutralino. As pointed
out in this paper, the heaviness of
the dark matter
particles help identify the signals from the background.

\vspace{1cm}

\noindent{\bf Acknowledgment:}

This work is partly supported by the National Natural Science Foundation
of China. We thank T. Han, S. Nussinov and R. Peccei for discussions.
We thank R. Cousins for bringing Ref\cite{fl} into our
attention during the final stage of this work.

\end{document}